\documentclass{aa}
\usepackage{graphicx}
\usepackage{psfig}
\usepackage{epsfig}
\begin{document}
\title{Circumstellar dust shells of hot post-AGB stars
\thanks{Based on observations with ISO, an ESA project with instruments 
funded by ESA Member States (especially the PI countries: France,
Germany, the Netherlands and the United Kingdom) and with the participation of 
ISAS and NASA.}}
\author{G. Gauba 
\and M. Parthasarathy} 
\institute{Indian Institute of Astrophysics, Koramangala, Bangalore 560034, 
India}
\offprints{G. Gauba,
\email{gsarkar@iucaa.ernet.in}}
\date{Received / Accepted}
\authorrunning{G.Gauba and M.Parthasarathy}
\titlerunning{Dust shells of hot post-AGB stars}

\abstract{
Using a radiative transfer code (DUSTY) parameters of the circumstellar 
dust shells of 15 hot post-AGB stars have been derived. Combining the 
optical, near and far-infrared (ISO, IRAS) data of the stars, we have
reconstructed their spectral energy distributions (SEDs) and estimated 
the dust temperatures, mass loss rates, angular radii of the inner
boundary of the dust envelopes and the distances to these stars. 
The mass loss rates (10$^{-6}-10^{-5}$M$_{\odot}$yr$^{-1}$) 
are intermediate between stars at the tip of the AGB and the PN phase. 
We have also studied the ISO spectra of 7 of these stars. Amorphous and 
crystalline silicate features were observed in IRAS14331-6435 (Hen3-1013), 
IRAS18062+2410 (SAO85766) and IRAS22023+5249 (LSIII +5224) indicating 
oxygen-rich circumstellar dust shells. The presence of unidentified 
infrared (UIR) band at 7.7$\mu$, SiC emission at 11.5$\mu$ and the "26$\mu$"
and "main 30$\mu$" features in the ISO spectrum of IRAS17311-4924 (Hen3-1428) 
suggest that the central star may be carbon-rich. The ISO spectrum of 
IRAS17423-1755 (Hen3-1475) shows a broad absorption feature at 3.1$\mu$ due 
to C$_{2}$H$_{2}$ and/or HCN which is usually detected in the circumstellar 
shells of carbon-rich stars.
\keywords{Stars: AGB and post-AGB --- Stars: early-type ---
Stars: evolution --- Stars: circumstellar matter --- Infrared: stars}
}
\maketitle

\section{Introduction}

In the evolution of low and intermediate mass stars (0.8 $-$ 8M$_{\odot}$),
the post-asymptotic giant branch (post-AGB) or protoplanetary nebula (PPN) phase 
is a transition stage from the tip of the AGB to the planetary nebula (PN) 
stage (Kwok, 1993). The hot post-AGB stars form an evolutionary link between 
the cooler G,F,A supergiant post-AGB stars (Parthasarathy \& Pottasch, 1986) 
and the hotter O-B central stars of PNe (Parthasarathy, 1993a). 

Analysis of the UV(IUE) spectra of hot post-AGB stars
(Gauba \& Parthasarathy, 2003), revealed that in many cases, 
the hot (OB) central stars of PPNe are partially obscured by 
circumstellar dust shells. Stars on the AGB and beyond are characterised 
by severe mass loss (10$^{-8}$ $-$ 10$^{-3}$ M$_{\odot}$ yr$^{-1}$) which 
results in the formation of circumstellar envelopes. The physical mechanisms 
responsible for the intensive mass loss from AGB stars are not well understood 
although the most promising mechanism to date involves radiation pressure on the
dust grains (Tielens, 1983). While AGB stars appear to have spherically
symmetric dust outflows (eg. Habing \& Blommaert, 1993), PN tend to
have axially symmetric inner regions and spherical outer halos
(eg. Schwarz et al., 1992). Inorder to understand the mass loss mechanisms, 
wind velocities and time scales responsible for the evolution of PNe,
we need to study the circumstellar environment of the stage intermediate between 
the AGB and the PN phase, i.e. the post-AGB/PPN phase. Circumstellar dust
shells of some cooler post-AGB stars (eg. Hoogzaad et al., 2002; Hony 
et al., 2003) and PNe (eg. Siebenmorgen et al., 1994) have been modelled
to derive the dust composition, mass loss rates and dynamical ages.  

As a consequence of dredge-up of byproducts of helium burning to the 
surface of stars on the AGB, the oxygen-rich atmospheres of some of these stars may be 
transformed into carbon-rich atmospheres (see eg.  Iben \& Renzini, 1983). 
This change of chemistry would also be reflected in the composition of the dust 
grains formed in the cirumstellar envelopes of AGB and post-AGB stars.
With the resolution and wavelength coverage of the ISO mission 
(Kessler et al.,1996) the detection of prominent gas and solid state features 
specific to oxygen-rich and carbon-rich chemistries became possible. 
Amorphous and crystalline silicate features and crystalline water have been 
reported in the ISO spectra of some AGB and post-AGB stars and 
the nebulae surrounding [WC] central stars of PNe (see eg. Waters \& Molster, 1999;
Hoogzaad et al., 2002). Hrivnak et al. (2000) detected the "21$\mu$" and "30$\mu$"
emission features besides the unidentified infrared (UIR) emission bands 
at 3.3, 6.2, 7.7 and 11.3$\mu$ in the ISO spectra of a sample of carbon-rich PPNe. 

Bogdanov (2000, 2002, 2003) modelled the complete spectral energy distribution (SED) of
three hot post-AGB stars, IRAS18062+2410 (SAO85766), IRAS19590-1249 (LSIV-12 111) and
IRAS20462+3416 (LSII+34 26) using radiative transfer codes and derived their mass loss rates, 
inner radii of the dust envelopes, optical depth of the envelopes and the distances 
to these stars. We need to study a bigger sample of such stars to understand the 
evolution of the infrared spectrum as the stars evolve from the cooler post-AGB phase 
to the hot central stars of PNe. In this paper, we have used the radiative transfer 
code, DUSTY (Ivezi\'c et al., 1999) to model the circumstellar dust shells of 15 hot 
post-AGB stars. Additionally, 7 stars from our list were found to have ISO spectra. We 
also report the analysis of the ISO spectra of these stars.

\section{Target Selection}

The hot post-AGB stars for this study (Table 1) were selected from the
papers of Parthasarathy \& Pottasch 1989, Parthasarathy, 1993b and 
Parthasarathy et al., 2000a. High and low resolution optical spectra 
have confirmed the post-AGB nature of several of these stars, 
eg. IRAS17119-5926 (Hen3-1357; Parthasarathy et al., 1993c, 1995; Bobrowsky et 
al., 1998), IRAS18062+2410 (SAO 85766; Parthasarathy et al., 2000b; Mooney et al., 2002;
Ryans et al., 2003), IRAS19590-1249 (LSIV-12 111; McCausland et al., 1992;
Conlon et al., 1993a, 1993b; Mooney et al., 2002; Ryans et al., 2003)
and IRAS20462+3416 (LSII+34 26; Parthasarathy, 1993a; Garc\'ia-Lario et al., 1997a).
The UV(IUE) spectra of some hot post-AGB stars listed in Table 1 showed violet 
shifted P-Cygni profiles of CIV and NV indicating stellar wind and 
post-AGB mass loss (Gauba \& Parthasarathy, 2003). 

\section{ISO observations}

The ISO data archive was searched for spectra of the hot post-AGB
stars listed in Table 1. Six of the fifteen sources were found to have 
only SWS (Short Wavelength Spectrometer) spectra 
while one source, IRAS 17423-1755 (Hen3-1475) had both SWS and LWS 
(Long Wavelength Spectrometer) spectra. Log of the observations is 
given in Table 2. ISO SWS spectra have a wavelength coverage from 
2.38 $-$ 45.2$\mu$. The LWS spectra extend from 43 $-$ 197$\mu$. 

Our objects were observed in the low resolution (AOT 01) mode of
the SWS instrument (de Graauw et al., 1996). A spectrum scanned 
with SWS contains 12 subspectra, that each consist of two scans, 
one in the direction of decreasing wavelength ('up' scan) and the 
other in the direction of increasing wavelength ('down' scan). There 
are small regions of overlap in the wavelength between the sub-spectra. 
Each sub-spectrum is recorded by 12 independent detectors.  

The LWS observations were carried out in LWS01 mode, covering the 
full spectral range at a resolution ($\lambda$/$\Delta \lambda$)
of $\sim$ 200. The characteristics of the ISO LWS instrument are 
described in Clegg et al. (1996) and the calibration of the instrument
is described in Swinyard et al. (1996).

\vspace{-0.22cm}

\section{Analysis}

In this section we describe the analysis of the ISO spectra 
and the modelling of the SEDs of the hot post-AGB stars.

\subsection{ISO Data Analysis}

Offline processed ISO SWS01 (OLP version 10.1) and LWS01 
(OLP version 10) data were retrieved from the ISO data archive. These 
were further processed using ISAP (ISO Spectroscopic Analysis Package) 
version 2.1. 

\subsubsection{SWS}

The data analysis using ISAP consisted of extensive bad data removal 
primarily to minimize the effect of cosmic ray hits. All detectors 
were compared to identify possible features. For each sub-band, the 
flux level of the 12 detectors were shifted and brought to a mean value. 
They were then averaged, using median clipping to discard points that 
lay more than 2.5$\sigma$ from the median flux. The spectra were 
averaged typically to a resolution of of 300, 500, 800 and 1500 
for SWS01 data taken with speed 1, 2, 3 and 4 respectively (Table 2). 
Appropriate scaling factors were applied to the averaged spectra of 
each sub-band to form a continous spectrum from 2.38 $-$ 45.2$\mu$. 
The data of subband 3E (27.5$-$29.0 $\mu$) are generally noisy 
and unreliable (see eg. Hrivnak et al., 2000, Hony et al., 2002).   
Figs. 1a, b, c and 2a show the SWS spectra of the hot post-AGB
stars. The SWS data of IRAS14331-6435 (Hen3-1013), IRAS18062+2410
(SAO 85766) and IRAS22023+5249 (LSIII +5224) below 7$\mu$ and 
that of IRAS16206-5956 (SAO 243756), IRAS22495+5134 (LSIII +5142) 
below 12.5$\mu$ were noisy and have been excluded. This is also 
evident from the low IRAS 12$\mu$ flux ($<$ 5 Jy) of these stars. 
Identification of the infrared spectral features is based on Waters \& Molster 
(1999), Cox (1993), Hrivnak et al. (2000), Volk et al. (2002), 
Cernicharo et al. (1999) and J$\o$rgensen et al. (2000).

\subsubsection{LWS}

Reduction of the LWS observation of IRAS17423-1755 (Hen3-1475) consisted of 
extensive bad data removal using ISAP rebinning on a fixed resolution grid of 
$\lambda$/$\Delta \lambda$ = 250. Appropriate scaling factors were applied to 
the data from different LWS detectors to form a continuous spectrum  (Fig. 2c).
The LWS spectrum of IRAS17423-1755 (Hen3-1475) appears featureless. 

\subsubsection{Joining the SWS and LWS spectra}

Although the spectral shape is very reliable, the absolute flux 
calibration uncertainity is 30\% for the SWS at 45$\mu$ 
(Schaeidt et al., 1996) and 10$-$15\% for the LWS at 
45$\mu$ (Swinyard et al., 1998). The SWS and LWS spectra 
of IRAS17423-1755 (Hen3-1475) were scaled according to their 
fluxes in the overlap region. The difference between the flux 
levels of LWS and SWS in the overlap region was smaller than 
30\% which is acceptable within the limits of the combined 
error bars. The combined SWS-LWS spectra of
IRAS17423-1755 (Hen3-1475) was used in Fig. 3. 

\subsection{Spectral energy distributions (SEDs)}

To re-construct the spectral energy distributions (SEDs) of the 
objects, we combined the ISO data with available U,B,V,R,I,J,H,K,L,L' 
and M magnitudes of the stars from literature (Table 3a). We also 
searched the 2MASS (2Micron All Sky Survey) Catalog within 15\arcsec 
of each object for their JHK magnitudes and the Midcourse Space Experiment 
(MSX) catalog within 3\arcsec of each object. The 2MASS data was 
included only when it was found to be free from confusion flags. MSX fluxes were 
found only for IRAS17460-3114 (SAO 209306; Table 3b). Infrared data 
(8.7, 10, 11.4, 12.6 and 19.5$\mu$) on IRAS18062+2410 
(SAO 85766; Table 3c) was obtained from Lawrence et al. (1990). 

\subsubsection{Central star temperatures} 

The temperatures of the central stars (Table 4) are mainly from Gauba 
\& Parthasarathy (2003). For IRAS12584-4837 (Hen3-847)
and IRAS17423-1755 (Hen3-1475), we estimated the central star 
temperatures based on their spectral types. For the PN, IRAS22495+5134 (LSIII +5142)
we used a central star temperature of 35000K (Tylenda \& Stasi\'nska, 1994).
Gauba \& Parthasarathy (2003) found that the UV(IUE)
spectrum of IRAS22023+5249 (LSIII +5224) closely resembles that of
a B2-supergiant. Hence we adopted a temperature
of 18500K for the star, corresponding to the spectral type B2I.
LTE analysis of the high resolution optical spectra of
IRAS18062+2410(SAO85766) was carried out by Mooney et al. (2002) and
Parthasarathy et al. (2000b). Recently, using non-LTE analysis, Ryans et al. (2003) 
reported an effective temperature of 20750K for IRAS18062+2410 (SAO85766) and 20500K
for IRAS19590-1249 (LSIV-12 111). In this paper, we have adopted 
the temperature estimates by Ryans et al. (2003). 

\subsubsection{Reddening}

The interstellar extinction (E(B$-$V)$_{\rm I.S.}$) in the direction 
of the stars were estimated using the Diffuse Infrared Background 
Experiment (DIRBE)/IRAS dust maps (Schlegel et al., 1998; Table 4).
The optical spectral types of the stars are mainly from Parthasarathy
et al. (2000a). The intrinsic B$-$V values, (B$-$V)$_{\rm o}$, for
the optical spectral types of the stars, were taken from 
Schmidt-Kaler (1982). We adopted (B$-$V)$_{\rm o}$=$-$0.20 for 
IRAS12584-4837 (Hen3-847) and IRAS17423-1755 (Hen3-1475)
corresponding to T$_{\rm eff}$=20000K. For IRAS22023+5249 (LSIII +5224), we 
used (B$-$V)$_{\rm o}$=$-$0.16 corresponding to B2I spectral type 
and for the PN, IRAS22495+5134 (LSIII +5142), we adopted 
(B$-$V)$_{\rm o}$=$-$0.30 correspoding to T$_{\rm eff}$=35000K.
Using the observed and intrinsic B$-$V values
we derived the total (interstellar plus circumstellar) extinction,
E(B$-$V)$_{\rm total}$ (=(B$-$V)$_{\rm obs}$ $-$ (B$-$V)$_{\rm o}$)
towards these stars. Comparing E(B$-$V)$_{\rm total}$ and 
E(B$-$V)$_{\rm I.S.}$, it is evident that there is considerable 
circumstellar extinction in most cases. Gauba \& Parthasarathy (2003) 
found that the circumstellar extinction law in the UV 
(from $\sim$ 1300\AA~ to 3200\AA~) varies linearly as $\lambda^{-1}$ 
in the case of IRAS13266-5551 (CPD-55 5588), IRAS14331-6435 (Hen3-1013),
IRAS16206-5956 (SAO 243756), IRAS17311-4924 (Hen3-1428), 
IRAS18023-3409 (LSS 4634), IRAS18062+2410 (SAO 85766), 
IRAS18371-3159 (LSE 63), IRAS22023+5249 (LSIII +5224) and 
IRAS22495+5134 (LSIII +5142)

Since little is known  about circumstellar extinction laws, 
we corrected the observed optical and near infrared magnitudes 
of the stars for the total extinction (E(B$-$V)$_{\rm total}$) 
using the standard extinction laws by Rieke \& Lebofsky (1985).
In particular, Rieke \& Lebofsky (1985) assume R$_{v}$=3.1, 
where, R$_{v}$=A$_{v}$/E(B$-$V). Although this is true for 
interstellar extinction, it may not be strictly true 
for circumstellar extinction obeying a $\lambda^{-1}$ law
in the UV. In particular, R$_{v}$ may be different from 
3.1 in the case of circumstellar extinction. 

\subsubsection{Modelling the circumstellar dust shells with DUSTY code}

The use of the radiative transfer code, DUSTY (Ivezi\'c et al., 1999)
for modelling the circumstellar dust shells of hot post-AGB stars was
described in Gauba et al. (2003). DUSTY uses six different grain
types : 'warm' (Sil-Ow) and 'cold' (Sil-Oc) silicates from Ossenkopf 
et al. (1992), silicates and graphites (Sil-Dl and grf-DL) from 
Draine and Lee (1984), amorphous carbon (amC-Hn) from Hanner (1988)
and SiC (SiC-Pg) from P\'egouri\'e (1988). The central stars were
assumed to be point sources at the centers of the spherical density
distributions. The SEDs of the central stars were assumed to be 
Planckian. The standard Mathis, Rumpl, Nordsieck (MRN) 
(Mathis et al., 1977) power-law was used for the grain size (n(a)) 
distributions, i.e.  n(a) $\propto$ a$^{-q}$ for a$_{min}$ $\le$ a $\le$ a$_{max}$
with q=3.5, a(min)=0.005$\mu$ and a(max)=0.25$\mu$. For each object, the 
dust temperature (T$_{d}$) on the inner shell boundary and the optical 
depth ($\tau$) at 0.55$\mu$ (V-band) were varied. We assumed an inverse 
square law (y$^{-2}$) for the spherical density distribution. The shell 
was assumed to extend to 1000 times its inner radius. We adopted the 
fits for which the sum of squares of the deviations between the observed 
and modelled fluxes (after scaling) were a minimum.  Table 4 lists 
the adopted input parameters. Fig. 3 shows the spectral energy 
distribution of the stars. DUSTY does not allow simultaneous modelling 
of warm and cold dust shells. Hence, the cold dust in the case of 
IRAS12584-4837 (Hen3-847) and IRAS17423-1755 (Hen3-1475) had to be 
modelled and treated independent of the warm dust around these stars. 

Having fixed T$_{d}$ and $\tau$, we then used the gas-dynamical mode 
of the DUSTY code to derive the inner radii, r1(cm) where the dust temperatures 
(T$_{d}$) are specified and the mass-loss rates ($\dot M$). 
The radius scales in proportion to L$^{1/2}$ where L is the luminosity
and the code output value corresponds to L=10$^{4}$L$_{\odot}$. The 
mass-loss rate scales in proportion to L$^{3/4}$(r$_{gd}\rho_{s})^{1/2}$ where,
the gas-to-dust mass ratio, r$_{gd}$=200 and the dust grain density,
$\rho_{s}$=3 g cm$^{-3}$. The hot post-AGB stars discussed in this paper 
have a range of core-masses (Gauba \& Parthasarathy, 2003). 
Pottasch (1992) pointed out that the white 
dwarf distribution is sharply peaked with a mean mass between 0.56 and 
0.58 M$_{\odot}$ and central stars of PNe have core-masses which show a 
peak at approximately 0.6M$_{\odot}$. We carried out calculations for 
our hot post-AGB stars with core masses of 0.565M$_{\odot}$ and 0.605M$_{\odot}$ 
corresponding to luminosities of 4500L$_{\odot}$ (Sch\"onberner, 1983) 
and 6300L$_{\odot}$ (Bl\"ocker, 1995) respectively. Distances (d) 
to the stars were derived using r1 and the ratio of the observed and 
modelled fluxes at 0.55$\mu$. $\theta$ (=r1/d) is the angular radii of 
the inner boundary of the cold circumstellar dust envelopes.

Tables 5a and b list the respective values for T$_{d}$, r1, d,
$\theta$ and $\dot M$. All calculations were 
carried out using the best fit parameters for the cold circumstellar 
dust shells.

\subsection{Notes on individual objects}

The hot post-AGB stars in this paper, except IRAS19590-1249 (LSIV-12 111), 
have been described in Gauba et al. (2003) and Gauba \& Parthasarathy (2003).
Here, we describe, the ISO spectra and dust shell characteristics of some of
these objects. 

\#IRAS 12584-4837 (Hen3-847)

It was found to be variable in the optical (Kazarovets et al., 2000;
de Winter et al., 2001).The Hipparcos magnitudes at maximum and
minimum are 10\fm52 and 10\fm70 respectively. Comparison of the 
J,H,K magnitudes of the star from Fouque et al. (1992) and the
2MASS catalog indicate infrared variability as well. The SED of this 
shows the presence of both warm and cold circumstellar dust. We have
modelled the warm dust based on the data of Fouque et al. (1992) (775K) and
the 2MASS catalog (700K). The presence of warm circumstellar dust may 
indicate ongoing post-AGB mass loss. The use of amorphous carbon 
grains to model the warm dust and silicate grains for the cold dust, 
indicates that during its evolution along the AGB, the central star 
may have evolved from an oxygen-rich to a carbon-rich star.

\#IRAS 14331-6435 (Hen3-1013)

The ISO SWS spectrum reveals the presence of amorphous 
(10.8$\mu$) and crystalline silicates and/or water (33.6$\mu$, 40.4$\mu$, 
43.1$\mu$) in the circumstellar environment of this star. Crystalline 
silicates have been detected in the dust shells around evolved 
oxygen-rich stars (see eg. Waters et al.,1996, Waters \& Molster, 1999). 
Waters et al. (1996) found that these emission features are more prominent 
for objects with cooler dust shells (T $<$ 300K). Emission from 
crystalline water has been reported in the 40$-$70$\mu$ spectrum of the 
Frosty Leo nebula and other cool oxygen-rich envelopes (Omont et al., 1990). 

\#IRAS 16206-5956 (SAO 243756)

The ISO spectrum of this star is noisy and no features could be identified. 
The continuum from 12.5$\mu$ to 45.2$\mu$ was used in addition to the 
IRAS fluxes to better constrain the model fit to the SED. 

\#IRAS 17311-4924 (Hen3-1428)

The broad "30 $\mu$ emission feature" was detected in several AGB stars and PN sources 
(Forrest et al., 1981; Cox, 1993). More recently it was also detected in 
carbon-rich PPNs possessing the 21$\mu$ emission feature (Omont et al., 1995).
Hony et al. (2002) detected this feature in the ISO SWS spectrum of 
IRAS 17311-4924 (Hen3-1428) and a large sample of carbon-rich AGB stars 
(C-stars), post-AGB stars and PNe. Substructure in the 30$\mu$ feature was 
recognised by Szczerba et al. (1999). In the spectra of several PPNs
and carbon stars, it was resolved into two components 
(Hrivnak et al., 2000; Volk et al., 2000, 2002), the "26 $\mu$" 
and "main 30 $\mu$" feature (Fig. 1b). Goebel \& Moseley (1985) first 
suggested that the feature is due to magnesium sulfide (MgS). 
Hony et al. (2002) too identified MgS as the carrier of the "26 $\mu$" 
and "main 30 $\mu$" emission features. However, the 30$\mu$ band is never 
seen in oxygen-rich sources (Forrest et al., 1979). Since the feature is seen 
only in carbon-rich objects, the suggestion that its carrier is a carbonaceous material 
continues to be appealing (Volk et al., 2002). We modelled the SED of 
IRAS 17311-4924 (Hen3-1428) using graphite and silicon carbide (SiC). 
However, we could not obtain a fit in the 30$\mu$ emission region 
(from $\sim$ 20 $-$ 30 $\mu$). 

The 11.5$\mu$ band (Fig. 1b) is attributed to SiC (Treffers \& Cohen, 1984)
and has been detected in carbon-rich evolved objects (see eg., Cernicharo et
al., 1989). Cox(1993) pointed out that all AGB stars where a 30$\mu$ emission band is present
show the 11.5$\mu$ band in emission with the notable exception of GL 3068 where
it is seen in absorption. However, the reverse is not true and some caron-rich
sources with a a prominent 11.5$\mu$ emission band do not show the 30$\mu$
band (V Cyg, S Cep and Y CVn). Using the DUSTY model, we
obtained a good fit to the SiC emission.

We also detected the 7.7 $\mu$ UIR feature in the ISO
spectrum of this object. UIR features are commonly attributed to 
polyaromatic hydrocarbons (PAHs) and have been detected in other 
carbon-rich protoplanetary nebulae (see eg. Beintema et al., 1996; Hrivnak et al., 2000).

\#IRAS 17423-1755 (Hen3-1475)

Gauba et al. (2003) modelled the SED of this star using DUSTY. They found a warm dust
component at 1500K in addition to the cold-dust at 100K indicating
ongoing post-AGB mass-loss. Plotting the ISO spectrum of the star (2.38$\mu$ $-$
171 $\mu$) alongwith the photometric data from Gauba et al (2003), near-IR data 
from Garc\'ia-Lario et al. (1997b) and IRAS fluxes and using the DUSTY code, we 
detected a second warm dust component at 1000K. 
 
The broad absorption feature at 3.1$\mu$ seen in the ISO spectrum is due 
to C$_{2}$H$_{2}$ and/or HCN (Ridgway et al., 1978; Cernicharo et al., 1999; 
J$\o$rgensen et al., 2000). This feature is observed in carbon stars 
(Merrill \& Stein, 1976; Noguchi et al., 1977; Groenewegen et al., 1994). 
The far-IR flux distribution of the star was modelled with amorphous carbon and 
silicon carbide confirming the carbon rich nature of the circumstellar 
dust shell. The LWS spectrum appears featureless. 

\#IRAS 18062+2410 (SAO 85766)

High resolution optical spectra (Parthasarathy et al., 2000b;
Mooney et al., 2002; Ryans et al., 2003) indicated the underabundance 
of carbon in this star. The ISO spectrum shows strong emission due
to amorphous silicates at 10.8$\mu$ and 17.6$\mu$ in conformity
with an oxygen-rich chemistry (i.e. C/O $<$ 1) for the  central
star. 

The J,H,K magnitudes from Garc\'ia-Lario et al. (1997b) and the 2MASS
catalog indicate that the star is variable in the near-infrared.
Optical variations of the star were detected by Arkhipova et al. (1999).
The K,L,M fluxes by Lawrence et al. (1990) lie above the modelled SED (Fig. 3).
In particular the K-band flux of Lawrence et al. (1990) lies above the K-band 
flux estimated by Garc\'ia-Lario et al. (1997b). If the K,L,M fluxes of the star 
by Lawrence et al. (1990) were not overestimated, the observed mismatch may be 
due to the variable nature of this object. Gauba \& Parthasarathy (2003) reported 
variable circumstellar extinction which in addition to stellar pulsations may be 
due to a dusty torus in motion around the hot central star. 

Bogdanov (2000) modelled the SED of this star using a different radiative transfer 
code. They derived T$_{d}$=410K. This value is much higher than the dust temperature 
(T$_{d}$=230K) derived by us from the model fit. Bogdanov (2000) had used the 
K,L,M fluxes by Lawrence et al. (1990) only. Besides they did not use the ISO 
spectrum of the star to constrain the SED. We believe that we have a better estimate 
of the physical parameters of the star especially since our model 
gives a very good fit to the IRAS fluxes and ISO spectrum of the star. 

\#IRAS 18379-1707 (LSS 5112)

The star was found to be variable in the near-infrared based on the J,H,K magnitudes
from Garc\'ia-Lario et al. (1997b) and the 2MASS catalog.

\#IRAS 19590-1249 (LSIV-12 111)

It was classified from low dispersion spectroscopy by Kilkenny \& Pauls (1990)
as having a spectral type around B0. Based on an analysis of its high
resolution optical spectra, McCausland et al. (1992) and Conlon et al.
(1993a, 1993b) concluded that its chemical composition and atmospheric
parameters are consistent with a post-AGB evolutionary state. The 
presence of nebular emission lines of [NII], [OII] and [SII] in its 
optical spectrum suggest that it may be in the early stages of PN 
formation. Study of the atmospheric parameters and abundance analysis 
of the star have also been carried out
by Mooney et al (2002) and more recently a non-LTE analysis by Ryans
et al. (2003). Based on these studies it has been confirmed that the 
central star of LSIV-12 111 shows severe carbon deficiency.  
Using silicate grain composition for the dust envelope, we obtained a 
good fit to the SED (Fig. 3). We estimated a dust temperature of 120K 
which is in good agreement with the estimate of Conlon et al. (1993b)
based on the 100$\mu$ IRAS flux and the color-color diagram of 
Jourdain de Muizon et al. (1990). Recently, Bogdanov (2003) modelled
the spectral energy distribution of this star using the DUSTY code.
Although they adopted different values for the extinction and the
effective temperature of the central star, we find their derived 
values for the dust temperature, distance and mass loss rate
to be in good agreement with our results. 

\#IRAS 22023+5249 (LSIII +5224)

Amorphous (10.8$\mu$) and crystalline (33.6$\mu$) silicate features 
were detected in the ISO spectrum of the dust shell surrounding this 
star indicating the oxygen-rich nature of the central star. 

\#IRAS 22495+5134 (LSIII +5142)

The ISO spectrum of this PN is noisy and spectral features could not be 
identified. The angular radii ($\theta$) derived using the DUSTY code 
(Tables 5a and b) are in good agreement with the angular diameter of
4$\arcsec$ reported by Tylenda \& Stasi\'nska (1994) and Acker et al. (1992)
respectively. From the Hipparcos and Tycho Catalogues (ESA 1997), the PN
has a parallax of 0.42 mas. This implies a distance of 2.38 kpc. 
Assuming a core mass (M$_{\rm c}$) of 0.605 M$_{\odot}$, we derived a
distance of 2.17 kpc to the PN. 

\section{Discussion and Conclusions}

We have modelled the circumstellar dust shells of 15 hot post-AGB stars
using the radiative transfer code, DUSTY and derived their dust temperatures,
distances to the stars, mass loss rates and angular radii of the inner boundary 
of the dust envelopes (Tables 5a and b). These stars have detached dust 
shells (as is evident from the SEDs, Fig. 3), OB-giant or supergiant spectra and cold 
dust between 100$-$315K, satisfying the observational properties of PPNe 
as defined by Kwok (1993, 2001). In addition to the cold dust, warm dust 
was detected in the case of IRAS12584-4837 (Hen3-847) and IRAS17423-1755 
(Hen3-1475) indicating ongoing mass loss. From the grain types used for the 
model fits, we may infer the chemical composition of the circumstellar dust 
shells. The use of both silicate and amorphous carbon grains to model the SEDs of 
IRAS12584-4837 (Hen3-847) and IRAS17460-3114 (SAO 209306) suggests that 
the central stars in these two cases may have undergone a recent change 
from an oxygen-rich to a carbon-rich chemistry. Such hot post-AGB stars may evolve 
into the [WC] central stars of PNe. Recently, Waters
et al. (1998) detected carbon-rich PAH features in the near-infrared and
crystalline silicates in the far-infrared ISO spectra of two PNe with
[WC] central stars, BD+30 3639 and He2-113.

Observational evidence (eg. Chu et al., 1991) suggests that three winds
are involved in stripping the outer envelope of the AGB star on its 
way to becoming a PN (Marten et al., 1993; Frank, 1994) : the
spherically symmetric AGB wind (eg. Habing \& Blommaert, 1993) when the
star loses mass at rates of $10^{-7}-10^{-6}$M$_{\odot}$yr$^{-1}$ with
a wind velocity of $\sim$ 10 kms$^{-1}$; the superwind phase when the
mass loss is thought to increase dramatically at the end of the AGB,
upto $10^{-5}-10^{-3}$M$_{\odot}$yr$^{-1}$, still with a wind velocity of
$\sim$ 10 kms$^{-1}$; once the superwind exhausts most of the AGB star's envelope, 
a fast wind with mass loss rate of $10^{-8}$M$_{\odot}$yr$^{-1}$ and
velocity of $\sim$ 1000 kms$^{-1}$ develops at some point during the PPN phase.
Velocities of 1000 kms$^{-1}$ and mass loss rates of $\sim$ $10^{-8}$M$_{\odot}$yr$^{-1}$
have been observed in the central stars of PNe (eg. Gauba et al., 2001).
For our hot post-AGB stars, we derived mass loss rates of 10$^{-5}-10^{-6}$M$_{\odot}$yr$^{-1}$. 
The mass loss rates ($\dot M$) scale with the 
gas-to-dust mass ratio (r$_{gd}$). We have adopted r$_{gd}$ = 200. For 
carbon-rich AGB and post-AGB stars values between 200 and 250 are often used 
(eg. Jura, 1986; Meixner et al., 1997). For the cool (F3Ib) post-AGB star, 
HD161796 (Parthasarathy \& Pottasch, 1986) with an oxygen-rich 
circumstellar environment, Hoogzaad et al. (2002) 
estimated r$_{gd}$ = 270.  Furthermore, our models assume that the dust density 
distribution falls off as y$^{-2}$ in the entire circumstellar dust shell. Such an 
assumption would break down in the case of episodic mass loss (Olofsson et al., 1990). 
Eg. episodic mass loss may have been responsible for the rapid evolution (30 $-$ 40 years) of 
IRAS17119-5926 (Hen3-1357) and IRAS18062+2410 (SAO85766) from B-type 
post-AGB supergiants to young PNe (Parthasarathy et al., 1993c, 1995; 
Bobrowsky et al., 1998, Parthasarathy et al., 2000b).

The proper motions ($\mu$) of the stars from the 
Tycho-2 Catalogue (Hog et al., 2000) have been listed in Tables 5a and b. 
Using the derived distances (d) in conjunction with the proper motions
we estimated the component of the stellar space velocities of the targets
tangent to the line of sight (V$_{\rm T}$). 
For IRAS17203-1534, IRAS18062+2410 (SAO85766) and IRAS18371-3159 
(LSE63), the large V$_{\rm T}$ values (Tables 5a and b) imply very high space 
velocities (V$_{\rm s}$ = (V$_{\rm T}^{2}$ $+$ V$_{\rm r}^{2}$)$^{1/2}$; where
V$_{\rm r}$ is the radial velocity of a star), close to the 
escape velocity from the Galaxy of 290 kms$^{-1}$ near the Sun.
Mooney et al. (2002) estimated a distance of 8.1 kpc to IRAS18062+2410 
(SAO85766) which is much greater than our estimate of 
$\sim$ 5 kpc. Such a large distance, if correct, would imply a still higher 
space velocity. We believe our distance estimates to be closer to
the actual values for these stars. However, the assumption
of spherical density distributions in our models, may be an over 
simplification for some of these objects. 
Eg. IRAS17423-1755 (Hen3-1475) has IRAS colors moderately close to
those of HD233517. HD233517 is unresolved and may have a disk 
instead of a spherical outflow (see eg. Jura, 2003, Fisher et al.,
2003). The predicted angular sizes of the inner radii of the dust shells
(Tables 5a and b) suggests that these objects should be easily 
resolvable in the mid-IR images with large ground based telescopes. 
Imaging of these objects in the IR would serve to test the basic assumptions 
such as those of spherical symmetry for our models.

In Table 6 we have compared the predicted and observed (V$-$J) values
for these stars ($\Delta$(V-J)=(V-J)$_{\rm predicted}$-(V-J)$_{\rm obs}$)
where, (V-J)$_{\rm predicted}$=(V-J)$_{\rm o}$ $+$ A$_{\rm V}$ $-$ A$_{\rm J}$. 
The intrinsic (V-J) colors ((V-J)$_{\rm o}$), for the spectral types of
the stars are from Ducati et al. (2001). For stars with emission lines in their
optical spectra, IRAS12584-4837 (Hen3-847), IRAS17423-1755 (Hen3-1475),  
IRAS22023+5249 (LSIII+5224)and for the PN, IRAS22495+5134 (LSIII+5142), 
(V-J)$_{o}$ could not be assumed. In the case of 
IRAS14331$-$6435 (Hen3-1013), IRAS17203-1534,
IRAS18062$+$2410 (SAO85766) and IRAS18379-1707 (LSS5112), we find 
significant differences between the values of (V-J)$_{\rm predicted}$ 
and (V-J)$_{\rm obs}$. On first sight, this would then raise a suspicion 
about the adopted E(B$-$V)$_{\rm total}$ values. However, we would like to point out
that the V and J magnitudes of these stars have not been recorded
simultaneously. Many of these stars are variable as evidenced from
the J,H,K data on IRAS12584-4837 (Hen3-847), IRAS18062+2410(SAO85766)
and IRAS18379-1707 (LSS5112). The V and J magnitudes may have be recorded
at different epochs of the variability cycle and hence it may
not be suitable to compare (V-J)$_{\rm predicted}$ and (V-J)$_{\rm obs}$.

We also studied the ISO spectra of 7 hot post-AGB stars, 
IRAS14331-6435 (Hen3-1013), IRAS16206-5956 (SAO243756), 
IRAS17311-4924 (Hen3-1428), IRAS17423-1755 (Hen3-1475), 
IRAS18062+2410 (SAO85766), IRAS22023+5249 (LSIII +5224) and IRAS22495+5134
(LSIII +5142). A weak amorphous silicate feature (10.8$\mu$) alongwith
crystalline silicate features was found in the dust shells of 
IRAS14331-6435 (Hen3-1013) and IRAS22023+5249 (LSIII +5224).
The 17.6$\mu$ amorphous silicate feature was missing in these two stars.
The post-AGB star IRAS18062+2410 (SAO85766) did not show evidence
for the presence of crystalline silicates but strong amorphous silicate
features at 10.8$\mu$ and 17.6$\mu$ were detected. Volk \& Kwok (1989)
predict that at dust temperatures of typically a few 100K for 
post-AGB stars, the spectrum should increase from 8
to 23$\mu$ and the 10.8$\mu$ and 17.6 $\mu$ silicate features should cease
to be observable. This appears to be consistent with the observed
spectral features and the dust temperatures of 230K, 130K and 120K 
for IRAS18062+2410 (SAO85766), IRAS14331-6435 (Hen3-1013) and
IRAS22023+5249 (LSIII +5224) respectively from our model fits. 
The presence of silicate features in these stars indicates the 
O-rich nature of the central stars. The formation of crystalline 
silicates in the circumstellar shells of post-AGB stars 
is still not well understood (see eg., Waters et al., 1996).
In contrast, PAH emission at 7.7$\mu$, the "26$\mu$" and "main 30$\mu$" features
and 11.5$\mu$ SiC emission in IRAS17311-4924 (Hen3-1428), are typical of 
circumstellar dust shells around carbon-rich post-AGB stars. However, the
21$\mu$ emission feature detected in several carbon-rich PPNe (Hrivnak et al., 2000) was 
notably absent in the ISO spectrum of IRAS17311-4924 (Hen3-1428).
Volk et al. (2002) pointed out that although all sources
with the 21$\mu$ emission feature also display the 
"26$\mu$" and "main 30$\mu$" features, the converse is not true. 
The hot post-AGB star, IRAS01005+7910 (Klochkova et al., 2002) which showed the "26"
and "main 30$\mu$" emission also did not show the 21$\mu$ emission feature
(Hrivnak et al., 2000). It may be that the dust grains responsible for 
the 21$\mu$ emission are destroyed as the central star evolves 
towards hotter temperatures. The broad absorption feature at 3.1$\mu$ in 
IRAS17423-1755 (Hen3-1475) attributed to C$_{2}$H$_{2}$ and/or HCN indicates 
that the central star may be carbon-rich. 

\acknowledgements
We are thankful to the referee for helpful comments. GG would also like 
to thank Luciano Rezzolla and John Miller of the Astrophysics sector at
SISSA, Trieste, Italy for providing computer facility during her stay in Italy.
The ISO Spectral Analysis Package (ISAP) is a joint development by 
the LWS and SWS Instrument Teams and Data Centers. Contributing institutes are 
CESR, IAS, IPAC, MPE, RAL and SRON.

\begin{table*}[h]
\begin{center}
\caption{Hot post-AGB stars}
\begin{tabular}{|c|c|c|c|c|c|c|c|c|c|}
\hline 

Star No. & IRAS & Name & Optical & l & b & \multicolumn{4}{c|}{IRAS Fluxes (Jy.)} \\ \cline{7-10} 
         &      &      & Sp. Type &  &   & 12$\mu$ & 25$\mu$ & 60$\mu$ & 100$\mu$ \\
\hline \hline

1. & 12584-4837 & Hen3-847     & Be$^{1}$ & 304.60&  +13.95& 36.07 & 48.75 & 13.04 & 3.31 \\
2. & 13266-5551 & -55 5588     & B1Ibe   & 308.30&  +6.36 & 0.76  & 35.90 & 35.43 & 11.66\\
3. & 14331-6435 & Hen3-1013    & B3Ie    & 313.89&  -4.20 & 4.04  & 108.70& 70.71 & 20.61\\
4. & 16206-5956 & SAO 243756   & A0Ia$^{2}$ & 326.77&  -7.49 & 0.36L & 11.04 & 12.30 & 4.83\\ 
5. & 17203-1534 &              & B1IIIpe & 8.55  &  +11.49& 0.32  & 10.70 & 6.88  & 3.37\\    
6. & 17311-4924 & Hen3-1428    & B1IIe   & 341.41&  -9.04 & 18.34 & 150.70& 58.74 & 17.78\\ 
7. & 17423-1755 & Hen3-1475    & Be      &  9.36&  +5.78 &  7.05 &  28.31& 63.68 & 33.43\\ 
8. & 17460-3114 &  SAO 209306  & O8III   & 358.42&  -1.88 & 6.26  & 20.82 & 12.20 &220.40L\\
9. & 18023-3409 & LSS 4634     & B2IIIe  & 357.61&  -6.31 & 0.26L & 2.94  & 1.82  & 25.64L\\ 
10.& 18062+2410 & SAO 85766    & B1I$^{3}$ & 50.67&  +19.79& 3.98  & 19.62 & 2.90 & 1.00L \\  
11.& 18371-3159 & LSE 63       & B1Iabe  & 2.92 & -11.82 & 0.25L &  6.31 &  5.16 & 1.95\\
12.& 18379-1707 & LSS 5112     & B1IIIpe & 16.50 & -5.42  & 1.67  & 23.76 &  7.12 & 3.66L\\
13.& 19590-1249 & LSIV-12 111  & B1Ibe   & 29.18 & -21.26 & 0.29: & 10.26 & 6.45 & 1.77: \\
14.& 22023+5249 & LSIII +5224  & B$^{4}$ & 99.30 &  -1.96 & 1.02  & 24.69 & 14.52 & 3.93L\\
15.& 22495+5134 & LSIII +5142  & PN$^{5}$ & 104.84&  -6.77 & 0.54  & 12.37 &  7.18 & 3.12\\ 

\hline
\end{tabular}

\vspace{0.2cm}

\noindent \parbox{16cm}{A colon {\bf :} indicates moderate quality IRAS flux,
{\bf L} is for an upper limit.\\
The spectral types are from Parthasarathy et al.
(2000a) except $^{1}$Kazarovets et al. (2000); $^{2}$Schild et al. (1983);
$^{3}$Parthasarathy et al. (2000b); $^{4}$Simbad database; $^{5}$Acker et al. (1992)}
\end{center}
\end{table*}

\begin{table*}
\begin{center}
\caption{Log of ISO observations}
\begin{tabular}{|c|c|c|c|c|c|c|} \hline
IRAS & Name &  Date of Obs. & Duration of Obs.(s) & TDT$^{a}$ & Mode$^{b}$ & Speed$^{c}$\\ \hline \hline
14331-6435 & Hen3-1013   &  14 July 1997 & 3454 & 60600607 & SWS01 & 3 \\
16206-5956 & SAO 243756  &  6 Sept. 1996 & 6538 & 29401311 & SWS01 & 4 \\
17311-4924 & Hen3-1428   &  28 Feb. 1996 & 1834 & 10300636 & SWS01 & 2 \\
17423-1755 & Hen3-1475   & 17 March 1997 & 1140 & 48700267 & SWS01 & 1 \\
           &             & 17 March 1997 &  800 & 48700168 & LWS01 & --\\
18062+2410 & SAO 85766   &  18 Feb. 1997 & 1140 & 46000275 & SWS01 & 1 \\
22023+5249 & LSIII +5224 &   5 Jan. 1997 & 1140 & 41600993 & SWS01 & 1 \\
22495+5134 & LSIII +5142 &   5 Jan. 1997 & 1140 & 41601295 & SWS01 & 1 \\

\hline
\end{tabular} 

\vspace{0.2cm}

\noindent \parbox{16cm}{$^{a}$TDT number uniquely identifies each ISO observation. 
$^{b}$SWS observing mode used (see de Graauw et al., 1996). $^{c}$Speed corresponds 
to the scan speed of observation.}
\end{center}
\end{table*}

\setcounter{table}{2}
\begin{table*}
\renewcommand{\thetable}{\arabic{table}a}
\begin{center}
\caption{Photometric data on hot post-AGB stars}
\begin{tabular}{|c|c|c|c|c|c|c|c|c|c|c|c|} \hline
IRAS &  U  &  B  & V   &  R  & I   & J   & H   & K   & L   & L$^{'}$ & M \\ 
     & mag & mag & mag & mag & mag & mag & mag & mag & mag & mag     & mag \\
\hline \hline 
12584-4837 &       & 10.65$^{a}$ & 10.58$^{a}$ &  &  & 9.99$^{b}$ & 9.14$^{b}$ & 7.49$^{b}$ & & 
4.53$^{b}$ & 3.59$^{b}$ \\
           &       &             &             &  &  & 10.18$^{e}$ & 9.42$^{e}$ & 7.80$^{e}$ & &
           &            \\      
13266-5551 & 10.33$^{c}$ & 10.99$^{c}$ & 10.68$^{c}$ &  &  & 9.96$^{d}$ & 9.84$^{d}$ & 9.70$^{d}$
& & & \\
14331-6435 & 11.11$^{c}$ & 11.48$^{c}$ & 10.90$^{c}$ &  &  & 9.35$^{d}$ & 9.03$^{d}$ & 8.72$^{d}$ 
& & & \\
16206-5956 &  9.87$^{c}$ & 10.07$^{c}$ & 9.76$^{c}$  &  &  & 9.00$^{e}$ & 8.87$^{e}$  & 8.71$^{e}$ & & & \\
17203-1534 &       & 12.37$^{a}$ & 12.02$^{a}$ &  &  & 10.95$^{e}$ & 10.81$^{e}$ & 10.66$^{e}$ 
& & & \\
17311-4924 & 10.52$^{f}$ & 11.08$^{f}$ & 10.68$^{f}$ &  &  & 9.74$^{d}$ & 9.54$^{d}$ & 9.19$^{d}$ 
& & & \\
17423-1755 & 13.57$^{g}$ & 13.3$^{h}$ & 12.64$^{g}$ & 11.75$^{g}$ & 10.91$^{g}$ & 9.61$^{d}$ & 8.32$^{d}$ & 6.80$^{d}$ & & & \\
17460-3114 & 7.45$^{c}$ & 8.17$^{c}$ & 7.94$^{c}$ &  &  & 7.32$^{e}$ & 7.31$^{e}$ & 7.26$^{e}$ 
& & & \\
18023-3409 & 11.70$^{c}$ & 12.01$^{c}$ & 11.55$^{c}$ & & & & & & & & \\
18062+2410 & 10.86$^{i}$ & 11.59$^{i}$ & 11.54$^{i}$ & & & 11.22$^{d}$ & 10.97$^{d}$ & 10.84$^{d}$;
10.21$^{j}$ & 9.61$^{j}$ & & 7.30$^{j}$ \\ 
           &             &             &             & & & 11.36$^{e}$ & 11.33$^{e}$ & 11.23$^{e}$
           &             & &            \\
18371-3159 & & 12.09$^{a}$ & 11.98$^{a}$ & & & & & & & & \\
18379-1707 & 11.94$^{c}$ & 12.38$^{c}$ & 11.93$^{c}$ & & & 10.76$^{d}$ & 10.55$^{d}$ & 10.33$^{d}$ 
& & & \\
           &             &             &             & & & 10.66$^{e}$ & 10.43$^{e}$ & 10.16$^{e}$
& & & \\
19590-1249 & 10.75$^{k}$ & 11.41$^{k}$ & 11.32$^{k}$ & & & 11.08$^{d}$ & 10.96$^{d}$ & 10.78$^{d}$ 
& & 10.34$^{l}$ & \\
22023+5249 &  & 13.21$^{a}$ & 12.52$^{a}$ & & & 11.30$^{d}$ & 11.11$^{d}$ & 10.83$^{d}$ & & & \\
22495+5134 &  & 12.00$^{a}$ & 11.78$^{a}$ &  &  & 11.82$^{e}$ & 11.76$^{e}$ & 11.57$^{e}$ & & & \\

\hline
\end{tabular}

\vspace{0.2cm}

\noindent \parbox{16cm}{Photometry is from : $^{a}$Hog et al.(2000); $^{b}$Fouque et al.(1992);
$^{c}$Reed(1998); $^{d}$Garc\'ia-Lario et al.(1997b); $^{e}$2MASS; $^{f}$Kozok (1985);
$^{g}$Gauba et al. (2003); $^{h}$Monet et al. (1998); $^{i}$Arkhipova et al. (1999); 
$^{j}$Lawrence et al.(1990); $^{k}$Arkhipova et al. (2002), $^{l}$Conlon et al. (1993a)} 

\end{center}
\end{table*}

\setcounter{table}{2}
\begin{table*}
\renewcommand{\thetable}{\arabic{table}b}
\begin{center}
\caption{MSX data}
\begin{tabular}{|c|c|c|c|c|} \hline
     & \multicolumn{4}{c|}{MSX Fluxes (Jansky)}\\ \cline{2-5}
IRAS & Band A & Band C & Band D & Band E \\ 
     & 8.28$\mu$ & 12.13$\mu$ & 14.65$\mu$ & 21.34$\mu$ \\ \hline \hline
17460-3114 & 4.0095 & 5.2726 & 6.9896 & 20.730 \\

\hline
\end{tabular}
\end{center}
\end{table*}

\setcounter{table}{2}
\begin{table*}
\renewcommand{\thetable}{\arabic{table}c}
\begin{center}
\caption{Infrared data on IRAS18062+2410 (SAO85766)}
\begin{tabular}{|c|c|c|c|c|c|} \hline
     & \multicolumn{5}{c|}{IR data in magnitudes} \\ \cline{2-6}
IRAS & 8.7$\mu$ & 10$\mu$ & 11.4$\mu$ & 12.6$\mu$ & 19.5$\mu$\\ \hline \hline
18062+2410 & 4.29 & 2.17 & 2.22 & 2.31 & $-$0.25\\

\hline
\end{tabular}
\end{center}
\end{table*}

\begin{table*}
\begin{center}
\caption{Input physical parameters for DUSTY and the adopted reddening values}
\begin{tabular}{|c|c|c|c|c|c|c|c|} \hline
IRAS & dust & E(B$-$V)    & E(B$-$V) & T$_{eff}$ & grain types $^{\dagger}$ & optical depth & T$_{d}$ \\
     & type & total       &  I.S.    & (K)     &                           & ($\tau$) at 0.55$\mu$ & (K)  \\ 
\hline \hline
12584-4837$^{*}$ & warm$^{1}$ & 0.27 & 0.18 & 20000 & amC-Hn & 0.26 & 775 \\
                 & warm$^{2}$ & 0.27 & 0.18 & 20000 & amC-Hn & 0.29 & 700 \\
           & cold & 0.27 & 0.18 & 20000 & Sil-DL \& SiC & 0.29 & 200 \\
13266-5551 & cold & 0.51 & 0.53 & 20800 & grf-DL & 0.12 & 160 \\
14331-6435 & cold & 0.71 &  --  & 16200 & Sil-Ow & 0.50 & 130 \\
16206-5956 & cold & 0.29 & 0.22 & 11200 & grf-DL \& amC-Hn & 0.16 & 170 \\
17203-1534 & cold & 0.61 & 0.44 & 19000 & Sil-Ow & 0.15 & 117 \\
17311-4924 & cold & 0.66 & 0.22 & 20300 & grf-DL \& SiC-Pg & 0.22 & 250 \\
17423-1755 & warm & 0.86 & 0.67 & 20000 & amC-Hn & 0.16 & 1500 \\
           & warm & 0.86 & 0.67 & 20000 & amC-Hn & 0.32 & 1000 \\    
           & cold & 0.86 & 0.67 & 20000 & SiC-Pg & 0.35 & 100 \\ 
17460-3114 & cold & 0.54 & --   & 35000 & Sil-DL \& grf-DL & 0.0015 & 315 \\
18023-3409 & cold & 0.70 & 0.44 & 20300 & grf-DL & 0.015 & 220 \\
18062+2410$^{*}$ & cold & 0.24 & 0.11 & 20750 & Sil-Oc \& Sil-DL & 0.32 & 230 \\
18371-3159 & cold & 0.30 & 0.15 & 20800 & grf-DL & 0.13 & 180 \\ 
18379-1707$^{*}$ & cold & 0.71 & --   & 19000 & Sil-DL & 0.13 & 140 \\
19590-1249 & cold & 0.29 & 0.20 & 20500 & Sil-Ow \& Sil-DL & 0.08 & 120 \\
22023+5249 & cold & 0.85 & --   & 18500 & Sil-Oc \& Sil-DL & 0.29 & 120 \\
22495+5134 & cold & 0.52 & 0.36 & 35000 & Sil-DL & 0.011 & 125 \\

\hline
\end{tabular}

\vspace{0.2cm}

\noindent \parbox{16cm}{$^{\dagger}$The grain types used for modelling the SEDs are
'warm' (Sil-Ow) and 'cold' (Sil-Oc) silicates from Ossenkopf et al., 1992;
silicates and graphites (Sil-Dl and grf-DL) from Draine and Lee (1984),
amorphous carbon (amC-Hn) from Hanner (1988) and SiC (SiC-Pg) from 
P\'egouri\'e (1988). $^{*}$IRAS12584-4837 (Hen3-847), IRAS18062+2410 (SAO85766) and 
IRAS18379-1707 (LSS5112) are variable in J,H,K bands (see Table 3a). Hence, two separate 
models satisfied the warm dust component in IRAS12584-4837 (Hen3-847). Model 1 conforms to the J,H,K data from
Fouque et al. (1992). Model 2 conforms to the J,H,K data from the 2MASS catalog. }
\end{center}
\end{table*}

\setcounter{table}{4}
\begin{table*}
\renewcommand{\thetable}{\arabic{table}a}
\begin{center}
\caption{Derived stellar and dust envelope parameters for M$_{c}$=0.565M$_{\odot}$}
\begin{tabular}{|c|c|c|c|c|c|c|c|} \hline
IRAS & r1   & d     & $\theta$ & $\dot M$ & $\mu$ & V$_{\rm T}$\\
     & (cm) & (kpc) & ($\arcsec$) & M$_{\odot}$yr$^{-1}$ & mas yr$^{-1}$ & km s$^{-1}$\\
\hline \hline
12584-4837 & 2.03X10$^{16}$ & 2.85 & 0.47 & 1.34X10$^{-5}$ &  9.96 & 134.89 \\
13266-5551 & 3.92X10$^{16}$ & 2.05 & 1.28 & 8.08X10$^{-6}$ &  8.22 &  79.52 \\
14331-6435 & 5.10X10$^{16}$ & 2.10 & 1.62 & 3.27X10$^{-5}$ &  6.75 &  66.66 \\
16206-5956 & 2.91X10$^{16}$ & 3.24 & 0.60 & 7.14X10$^{-6}$ &  3.96 &  60.55 \\
17203-1534 & 5.25X10$^{16}$ & 3.70 & 0.95 & 1.35X10$^{-5}$ &  9.90 & 173.80 \\
17311-4924 & 1.72X10$^{16}$ & 1.66 & 0.70 & 8.24X10$^{-6}$ &  4.75 &  37.52 \\
17423-1755 & 2.98X10$^{17}$ & 3.15 & 6.39 & 5.30X10$^{-5}$ &       &        \\ 
17460-3114 & 9.93X10$^{15}$ & 0.30 & 2.19 & 1.86X10$^{-7}$ &  3.90 &   5.55 \\
18023-3409 & 3.12X10$^{16}$ & 2.46 & 0.58 & 1.25X10$^{-6}$ & 13.93 & 161.04 \\
18062+2410 & 9.13X10$^{15}$ & 4.48 & 0.14 & 1.03X10$^{-5}$ & 12.59 & 271.26 \\
18371-3159 & 3.14X10$^{16}$ & 5.04 & 0.41 & 7.69X10$^{-6}$ &  7.94 & 188.18 \\
18379-1707 & 3.00X10$^{16}$ & 3.10 & 0.64 & 1.12X10$^{-5}$ &       &        \\
19590-1249 & 4.80X10$^{16}$ & 3.93 & 0.82 & 9.89X10$^{-6}$ &  7.56 & 143.34 \\
22023+5249 & 4.62X10$^{16}$ & 3.33 & 0.93 & 2.09X10$^{-5}$ &  4.96 &  78.37 \\
22495+5134 & 5.59X10$^{16}$ & 1.83 & 2.04 & 3.05X10$^{-6}$ &  2.97 &  25.60 \\

\hline
\end{tabular}
\end{center}
\end{table*}

\setcounter{table}{4}
\begin{table*}
\renewcommand{\thetable}{\arabic{table}b}
\begin{center}
\caption{Derived stellar and dust envelope parameters for M$_{c}$=0.605M$_{\odot}$}
\begin{tabular}{|c|c|c|c|c|c|c|c|} \hline
IRAS & r1   & d     & $\theta$ & $\dot M$ & $\mu$ & V$_{\rm T}$\\
     & (cm) & (kpc) & ($\arcsec$) & M$_{\odot}$yr$^{-1}$ & mas yr$^{-1}$ & km s$^{-1}$\\
\hline \hline
12584-4837 & 2.40X10$^{16}$ & 3.37 & 0.47 & 1.73X10$^{-5}$ &  9.96 & 157.37 \\
13266-5551 & 4.63X10$^{16}$ & 2.42 & 1.28 & 1.04X10$^{-5}$ &  8.22 &  95.03 \\
14331-6435 & 6.03X10$^{16}$ & 2.49 & 1.62 & 4.22X10$^{-5}$ &  6.75 &  80.00 \\
16206-5956 & 3.44X10$^{16}$ & 3.83 & 0.60 & 9.22X10$^{-6}$ &  3.96 &  72.19 \\
17203-1534 & 6.21X10$^{16}$ & 4.38 & 0.95 & 1.74X10$^{-5}$ &  9.90 & 204.03 \\
17311-4924 & 2.03X10$^{16}$ & 1.96 & 0.70 & 1.06X10$^{-5}$ &  4.75 &  44.15 \\
17423-1755 & 3.52X10$^{17}$ & 3.72 & 6.39 & 6.84X10$^{-5}$ &       &        \\
17460-3114 & 1.17X10$^{16}$ & 0.36 & 2.19 & 2.40X10$^{-7}$ &  3.90 &   6.65 \\
18023-3409 & 2.48X10$^{16}$ & 2.92 & 0.58 & 1.62X10$^{-6}$ & 13.93 & 194.20 \\
18062+2410 & 1.08X10$^{16}$ & 5.29 & 0.14 & 1.33X10$^{-5}$ & 12.59 & 314.09 \\
18371-3159 & 3.71X10$^{16}$ & 5.95 & 0.41 & 9.93X10$^{-6}$ &  7.94 & 221.38 \\
18379-1707 & 3.55X10$^{16}$ & 3.67 & 0.64 & 1.45X10$^{-5}$ &       &        \\
19590-1249 & 5.67X10$^{16}$ & 4.64 & 0.82 & 1.28X10$^{-5}$ &  7.56 & 170.64 \\
22023+5249 & 5.46X10$^{16}$ & 3.94 & 0.93 & 2.70X10$^{-5}$ &  4.96 &  94.04 \\
22495+5134 & 6.61X10$^{16}$ & 2.17 & 2.04 & 3.94X10$^{-6}$ &  2.97 &  30.60 \\

\hline
\end{tabular}
\end{center}
\end{table*}

\begin{table*}
\begin{center}
\caption{Predicted and observed (V-J) colors}
\begin{tabular}{|c|c|c|c|c|c|c|} \hline
IRAS & A$_{\rm V}$ = 3.1 X E(B-V)$_{\rm total}$ & A$_{\rm J}$ = 0.28 X A$_{\rm V}$ & 
(V-J)$_{\rm o}$ & (V-J)$_{\rm predicted}$ & (V-J)$_{\rm obs}$ & $\Delta$ (V-J) \\
\hline \hline

12584-4837 & 0.84 & 0.23 &   --    &   -- & 0.59$^{a}$ & -- \\
           &      &      &         &      & 0.40$^{b}$ & --  \\         
13266-5551 & 1.58 & 0.44 & $-$0.47 & 0.67 & 0.72     & $-$0.05 \\
14331-6435 & 2.20 & 0.62 & $-$0.37 & 1.21 & 1.55     & $-$0.34 \\
16206-5956 & 0.90 & 0.25 &    0.02 & 0.67 & --       & --  \\
17203-1534 & 1.89 & 0.53 & $-$0.47 & 0.89 & 1.07     & $-$0.18 \\ 
17311-4924 & 2.05 & 0.57 & $-$0.47 & 1.01 & 0.94     &    0.07 \\
17423-1755 & 2.67 & 0.75 &   --    &  --  & 3.03     &  --     \\
17460-3114 & 1.67 & 0.47 & $-$0.57 & 0.63 & 0.62     &    0.01 \\
18023-3409 & 2.17 & 0.61 & $-$0.47 &      & --       &  --     \\
18062+2410 & 0.74 & 0.21 & $-$0.47 & 0.06 & 0.32$^{c}$ & $-$0.26 \\        
           &      &      &         &      & 0.18$^{b}$ & $-$0.12 \\
18371-3159 & 0.93 & 0.26 & $-$0.47 & 0.20 & --       &  --     \\
18379-1707 & 2.20 & 0.62 & $-$0.47 & 1.11 & 1.17$^{c}$ & $-$0.06 \\
           &      &      &         &      & 1.27$^{b}$ & $-$1.16 \\
19590-1249 & 0.90 & 0.25 & $-$0.47 & 0.18 & 0.24     & $-$0.06 \\
22023+5249 & 2.63 & 0.74 &   --    &  --  & 1.22     &  --     \\
22495+5134 & 1.61 & 0.45 &   --    &  --  & $-$0.04  &  --     \\

\hline
\end{tabular}

\vspace{0.2cm}

\noindent \parbox{16cm}{Based on J,H,K data from $^{a}$Fouque et al. (1992);
$^{b}$2MASS and $^{c}$Garc\'ia-Lario et al. (1997b)}
\end{center}
\end{table*}

\setcounter{figure}{0}
\begin{figure*}
\renewcommand{\thefigure}{\arabic{figure}a}
\epsfig{figure=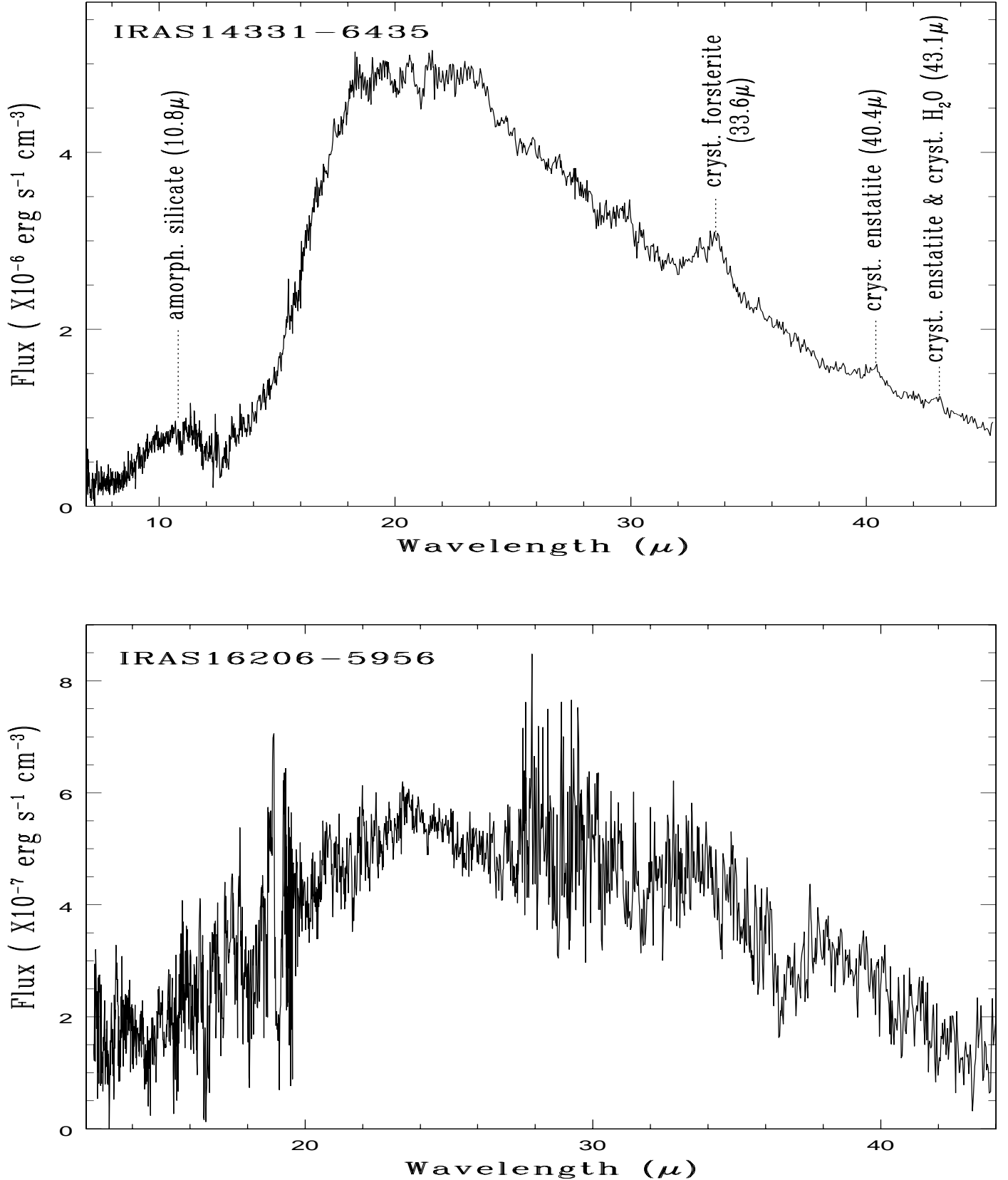, height=22cm, width=18.5cm}
\caption{The ISO SWS spectrum of IRAS14331-6435 (Hen3-1013) 
shows emission due to amorphous (10.8$\mu$) and crystalline 
silicates and/or H$_{2}$O (33.6$\mu$, 40.4$\mu$, 43.1$\mu$) 
(see eg. Waters \& Molster, 1999). ISO spectrum of IRAS16206-5956 
(SAO 243756) is noisy and only the continuum is seen here.}
\end{figure*}

\setcounter{figure}{0}
\begin{figure*}
\renewcommand{\thefigure}{\arabic{figure}b}
\epsfig{figure=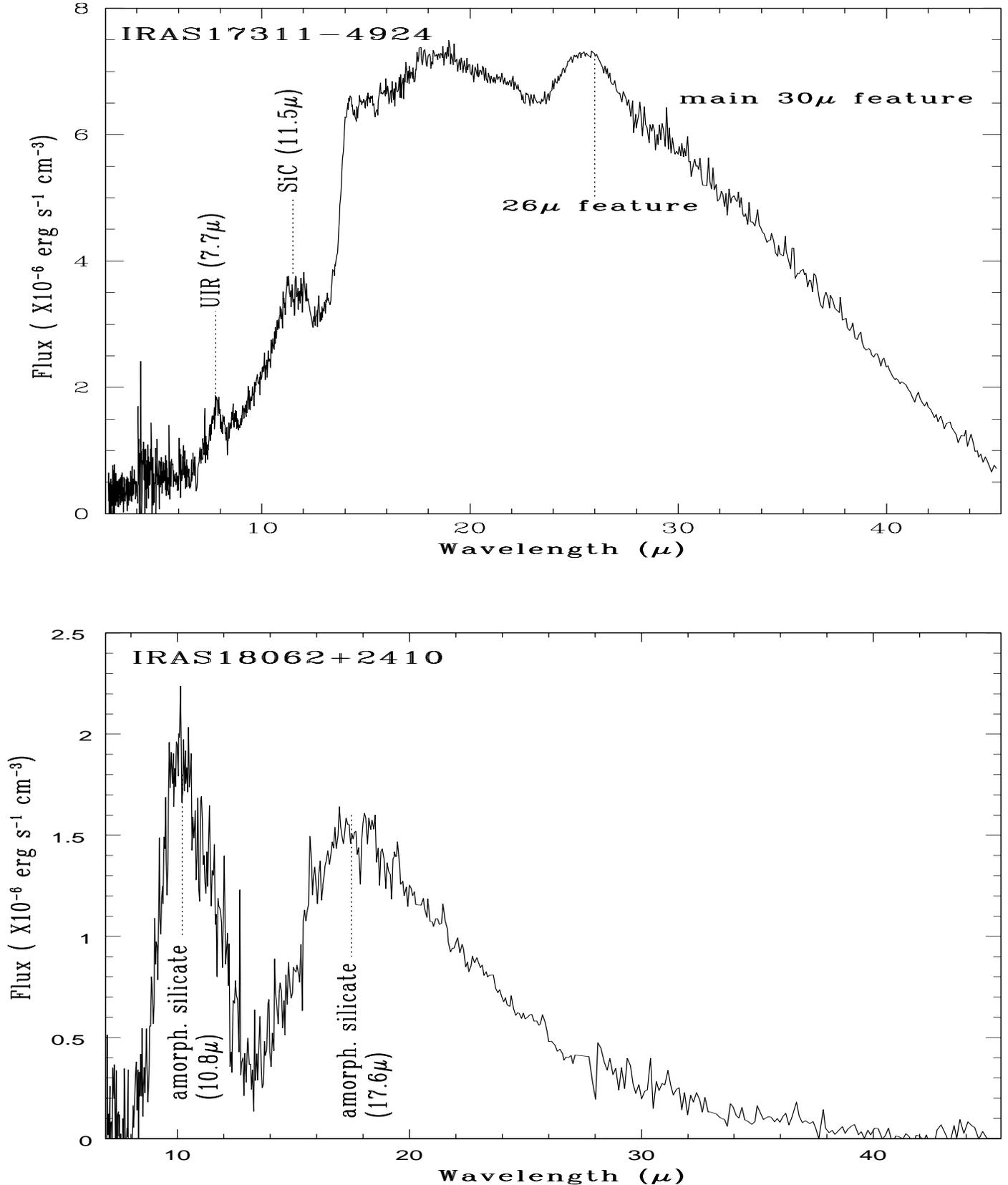, height=22cm, width=18.5cm}
\caption{The ISO SWS spectrum of IRAS17311-4924 (Hen3-1428) 
shows emission due to the UIR band at 7.7$\mu$,
SiC (11.5$\mu$) and the "30$\mu$ feature" (see eg. Hrivnak et al.,
2000; Volk et al., 2002). IRAS18062+2410 (SAO 85766)
shows emission due to amorphous silicates at 10.8$\mu$ and 17.6$\mu$
(see eg. Waters \& Molster, 1999).}

\end{figure*}

\setcounter{figure}{0}
\begin{figure*}
\renewcommand{\thefigure}{\arabic{figure}c}
\epsfig{figure=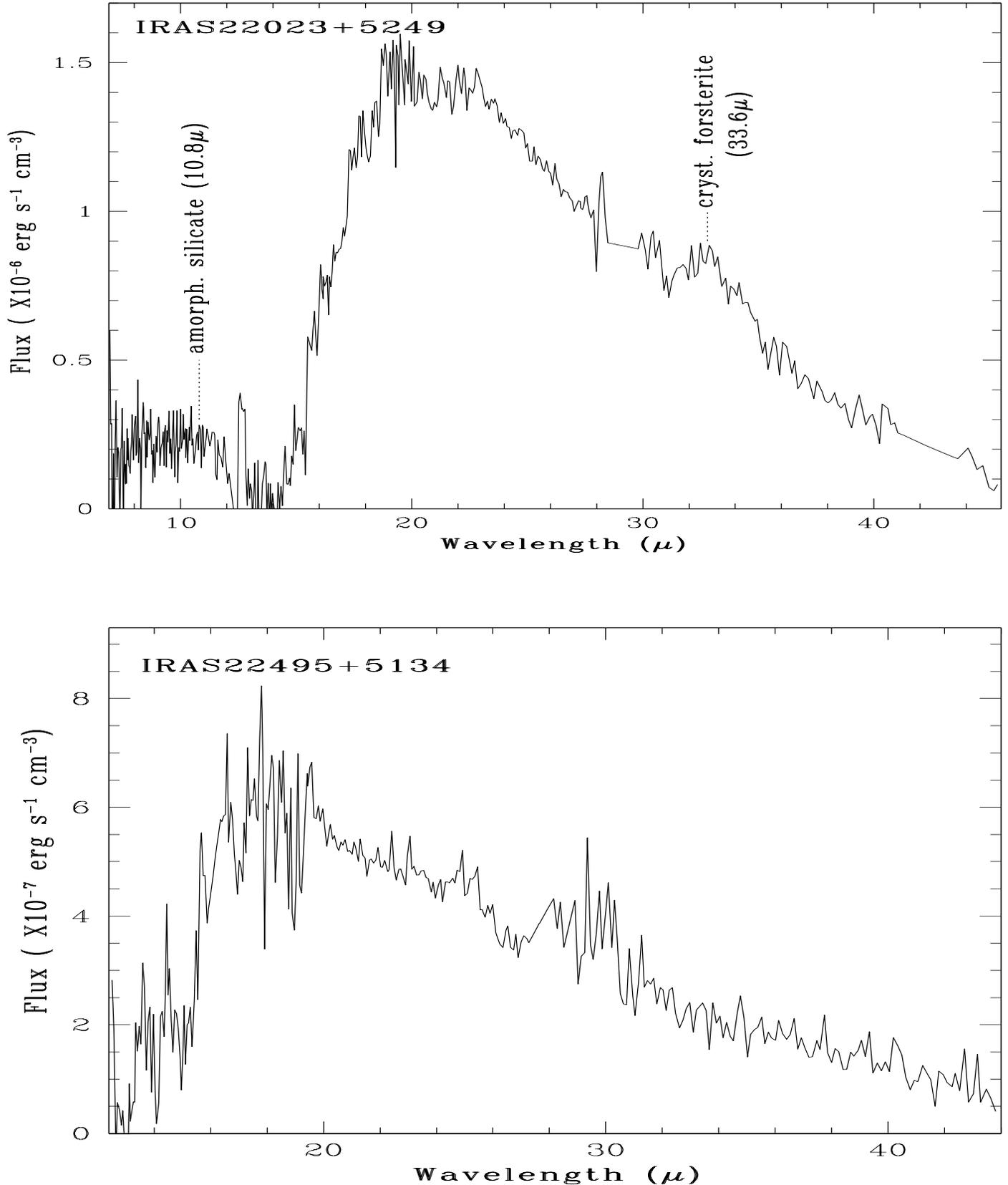, height=22cm, width=18.5cm}
\caption{The ISO SWS spectrum of IRAS22023+5249 (LSIII +5224)
shows emission due to amorphous (10.8$\mu$) and crystalline
(33.6$\mu$) silicates (see eg. Waters \& Molster, 1999). ISO SWS 
spectrum of IRAS22495+5134 (LSIII +5142) is noisy and only 
the continuum is seen here.}
\end{figure*}

\setcounter{figure}{1}
\begin{figure*}
\renewcommand{\thefigure}{\arabic{figure}}
\epsfig{figure=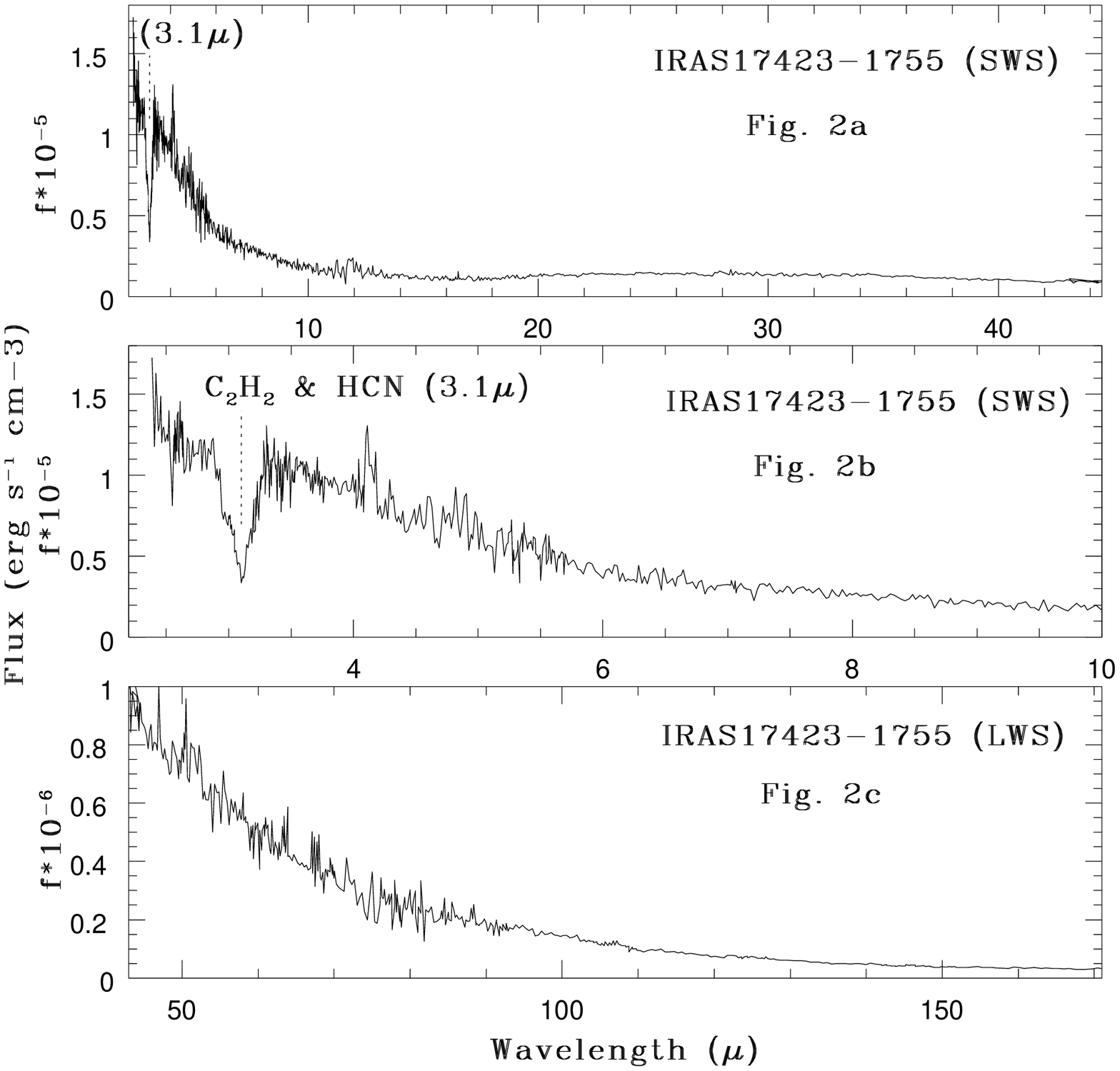, height=17cm, width=18.5cm}
\caption{ISO SWS (2.4$\mu$ $-$ 44.4$\mu$) and LWS 
(43$\mu$ $-$ 171$\mu$) spectra of IRAS17423-1755 (Hen3-1475) 
are presented in Figs. 2a and 2c respectively. The SWS spectrum 
shows a broad absorption feature at 3.1$\mu$ due to C$_{2}$H$_{2}$ 
and/or HCN (see eg. Cernicharo et al. 1999; J$\o$rgensen et al., 2000). 
This broad absorption feature is seen clearly in Fig. 2b showing 
the SWS spectrum of the star from 2.4$\mu$ $-$ 10$\mu$ only. 
The LWS spectrum of the star (Fig. 2c) appears featureless.} 
\end{figure*}

\setcounter{figure}{2}
\begin{figure*}
\epsfig{figure=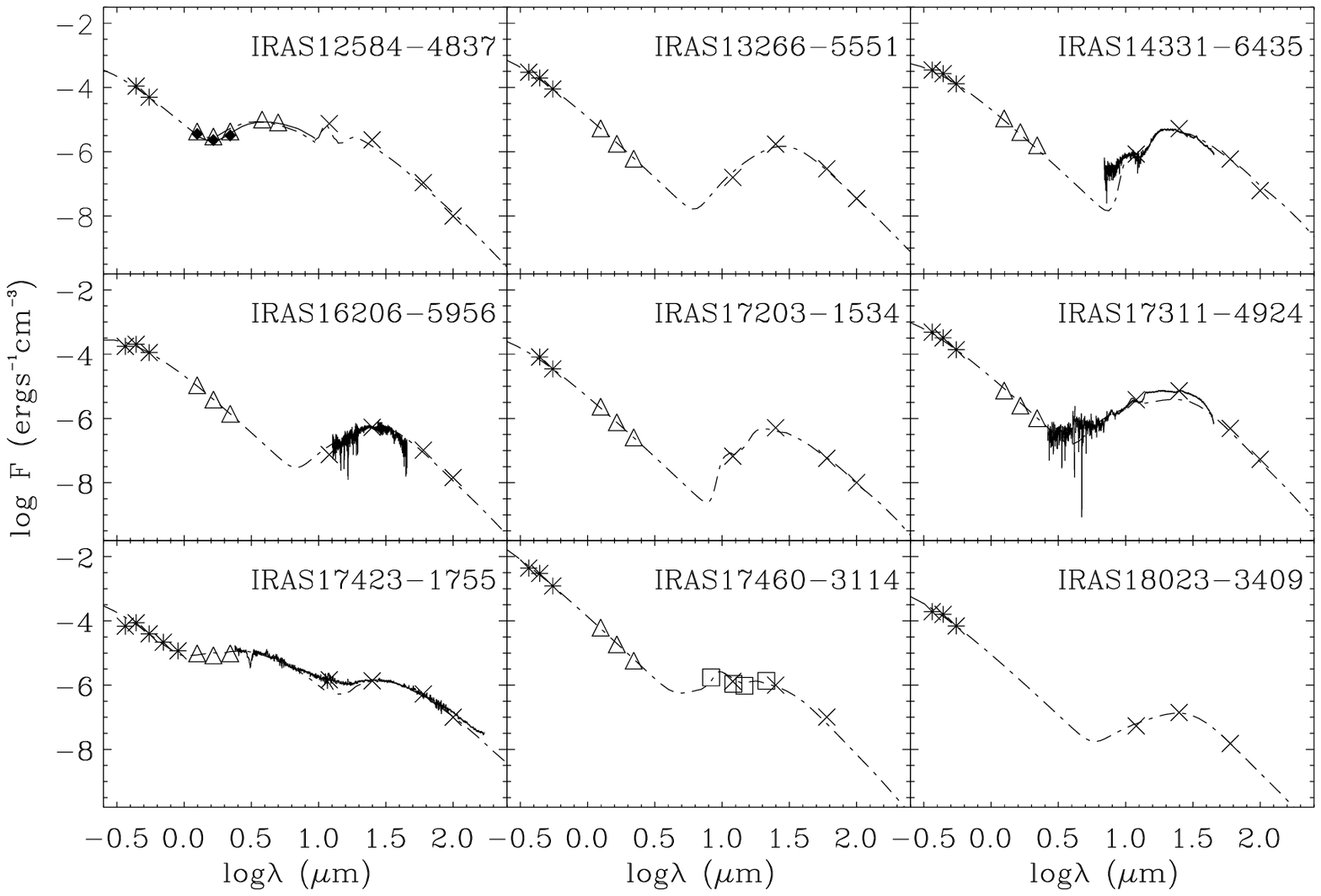, height=17.5cm, width=18.5cm}
\caption{Spectral energy distributions (SEDs) of the hot post-AGB stars.
UBVRI data (asterisk) are plotted alongwith JHK (triangle), MSX (square)
and IRAS data (cross). IRAS12584-4837 (Hen3-847), IRAS18062+2410 (SAO85766)
and IRAS18379-1707 (LSS5112) are variable in J,H and K bands. The
2MASS, J,H,K data of these three stars is represented by filled diamonds. 
ISO spectra of the hot post-AGB stars are plotted as solid lines. 
Dusty model fits are shown by dashed-dotted lines. The model fit corresponding
to the 2MASS J,H,K data of IRAS12584-4837 (Hen3-847) is shown by a solid line.
We could not obtain a fit to the "30$\mu$ emission feature" in 
IRAS17311-4924 (Hen3-1428). The IR (8.7, 10, 11.4, 12.6 and 19.5$\mu$) and K,L,M data of 
IRAS18062+2410 (SAO85766) by Lawrence et al. (1990) is indicated by diamonds. 
Notice the mismatch between the K-band flux of Lawrence et al. (1990; diamond) 
and Garc\'ia-Lario et al. (1997b; triangle) for the star. The L and M-band fluxes 
also do not lie on the modelled SED of the star (see Sec. 4.3).}
\end{figure*}

\setcounter{figure}{2}
\begin{figure*}
\epsfig{figure=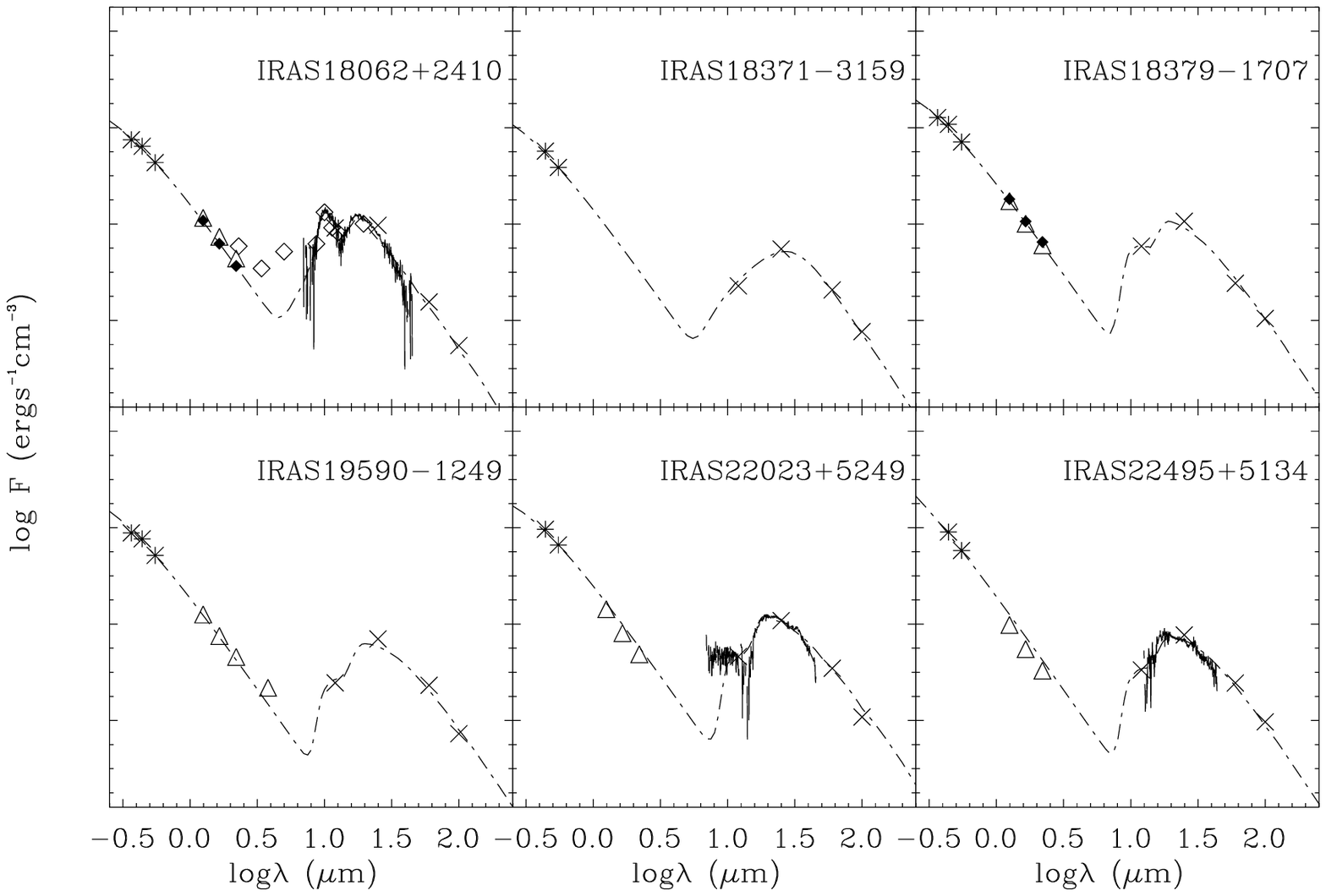, height=13.5cm, width=18.5cm}
\caption{contd....}
\end{figure*}

\end{document}